\RequirePackage{fix-cm}
\documentclass[twocolumn,epjc3]{svjour3}  
\smartqed  
\RequirePackage{graphicx}
 \RequirePackage{mathptmx}      
\RequirePackage{latexsym}
\RequirePackage[numbers,sort&compress]{natbib}
\RequirePackage[colorlinks,citecolor=blue,urlcolor=blue,linkcolor=blue]{hyperref}
\RequirePackage{amsmath,mathtools}

\journalname{Eur. Phys. J. A}
\begin{document}
\sloppy 
\title{Fluctuations and clustering of multiplicity in collisions of relativistic ions
}

\titlerunning{Fluctuations and clustering...}        

\author{Maciej Rybczy\'{n}ski\thanksref{e1,addr1} 
        \and
        Zbigniew W\l odarczyk\thanksref{addr1} 
}

\thankstext{e1}{e-mail: maciej.rybczynski@ujk.edu.pl}


\institute{Institute of Physics, Jan Kochanowski University, 25-406 Kielce, Poland \label{addr1}
}

\date{Received: date / Accepted: date}

\maketitle

\begin{abstract} We discuss the recently measured event-by-event multiplicity fluctuations in relativistic heavy-ion collisions. It is shown that the observed non-monotonic behaviour of the scaled variance of multiplicity distribution as a function of collision centrality (such effect is not observed in a widely used string-hadronic models of nuclear collisions) can be fully explained by the correlations between produced particles promoting cluster formation. We define a cluster as a quasi-neutral gas of charged and neutral particles which exhibits collective behaviour. The characteristic space scale of this shielding is the Debye length. Multiplicity distribution in a cluster is given by Negative Binomial distribution while the rest (reservoir), treated as a superposition of elementary collisions, is described by Binomial distribution. The ability to generate spatial structures (cluster phase) sign the propensity to self-organize of hadronic matter.

\keywords{fluctuations, multiplicity, multiparticle production}
\PACS{02.50.Ey, 05.10.Ln, 12.40.Ee}
\end{abstract}

\section{Introduction}
\label{sect:intro}

The studies of multiplicity fluctuations of particles produced in relativistic ion reactions are performed extensively since many years, because they may serve as a probe of dynamics present in particle production mechanism and possible creation of quark-gluon plasma. 

Collision of relativistic ions leads to a production of hot quark-gluon plasma, which cools and at $T=155\pm 10$~MeV~\cite{BraunMunzinger:2003zd,Bazavov:2014pvz,Bazavov:2018mes} transits to a hadron gas of that temperature. The hot quark-gluon system during the transition is effectively quenched by the cold physical vacuum. The so-called self-organized criticality is the appropriate mechnism leading to universal scale-free behavior~\cite{Castorina:2019pnb}. 

Self-organized criticality (SOC)~\cite{Bak:1987xua} is a property of non-equilibrium dynamical systems that have a critical point as an attractor (for review, see~\cite{jensen_soc,sornette_soc,pruessner_soc}). The macroscopic properties of such systems are characterized by the the spatial and/or temporal scale-invariance of the phase transition critical point. Unlike equilibrium systems which require the tuning of  parameters to enter a critical behavior, non-equilibrium SOC systems tune itself during evolution in the direction of criticality. A remarkable feature of active matter is the propensity to self-organize. One striking instance of this ability to generate spatial structures is the cluster phase, where cluster broadly distributed in size constantly move and evolve through particle exchange~\cite{Castorina:2019pnb}.

In the following sections we discuss imprints of multiplicity clustering on charged particles multiplicity fluctuations observed recently by the NA49 and NA61/SHINE experiments located at CERN SPS.

\section{Data on multiplicity fluctuations}
\label{sect:data}

In this work the multiplicity distribution $P\left(N\right)$ and its scaled variance $\omega$ are used to characterize
the multiplicity fluctuations. Let $P\left(N\right)$ denotes the probability to observe a particle multiplicity $N$ in a high energy nuclear
collision. By definition $P\left(N\right)$ is normalized to unity, $\sum_{N} P\left(N\right)=1$.
The scaled variance of multiplicity distribution, $\omega\left(N\right)$ is defined as:
\begin{equation}
\omega\left(N\right)=\frac{Var\left(N\right)}{\langle N\rangle}=\frac{\langle N^{2}\rangle - \langle N\rangle^{2}}{\langle N\rangle},
\label{eq:omega_def}
\end{equation}
where $Var\left(N\right)=\sum_{N}\left(N-\langle N\rangle\right)^{2}\cdot P\left(N\right)$ is the variance of the distribution and $\langle N\rangle=\sum_{N} N\cdot P\left(N\right)$ is the average multiplicity.

In many models the scaled variance of multiplicity distribution is independent of the number of particle production sources. Widely used models of nuclear collisions, the so-called superposition models, are based on the concept of particle emission from independent sources. The simplest example is the the Wounded Nucleon Model (WNM)~\cite{Bialas:1976ed}, in which the sources are wounded nucleons, i.e. the nucleons that have interacted at least once (usually calculated using Glauber model approach). In WNM, the scaled variance in nucleus-nucleus collisions is the same as in nucleon-nucleon interactions provided that the number of wounded nucleons is fixed. Also string-hadronic models predict similar values of $\omega$ for hadronic and nuclear collisions~\cite{Lungwitz:2007uc}. In a hadron-gas model~\cite{Begun:2006uu} the scaled variance of multiplicity distribution converges to a constant value with increasing volume of the system. In the special case of a hadron gas model, the so-called grand-canonical statistical formulation neglecting quantum effects and resonance decays multiplicity distribution is a Poisson (PD) one, namely:
\begin{equation}
P_{PD}\left(N\right)=\frac{\langle N\rangle^{N}}{N!}\cdot \exp\left(-\langle N\rangle\right). 
\label{eq:poiss_def}
\end{equation}
The variance of a PD is equal to its mean, and thus the scaled variance is $\omega=1$, independently of average multiplicity. It is then easy to find a possible discrepancy of the measured multiplicity distribution from the PD~\footnote{Notice that for Binomial distribution $\omega<1$ and for Negative Binomial distribution $\omega>1$.}. For a review, see Ref.~\cite{Heiselberg:2000fk}.

The NA49 and NA61/SHINE experiments located at CERN SPS analyzed multiplicity fluctuations of charged particles produced in p+p, Be+Be, Ar+Sc and Pb+Pb collisions~\cite{Alt:2006jr,Motornenko:2017klp,Grebieszkow:2017gqx}. Both experiments used scaled variance of multiplicity distribution, defined in Eq.~(\ref{eq:omega_def}), as a measure of multiplicity fluctuations. The NA49 Collaboration published data on multiplicity fluctuations in Pb+Pb reactions as a function of collision centrality~\cite{Alt:2006jr}. Unexpectedly, the measured scaled variance show very non-trivial centrality dependence. It is close to unity at completely central collisions but it manifests a prominent discrepancy from unity at peripheral interactions. The measurement has been performed at the collision center of mass energy $\sqrt{s_{NN}}=17.3$~GeV for particles produced in forward hemisphere in the restricted rapidity inverval $1.1 < y_{\pi} < 2.6$~\footnote{$y_{\pi}$ denotes rapidity calculated assuming mass of $\pi$ meson.} in the center of mass frame. The azimuthal acceptance has been also limited, and about 17\% of all produced charged particles have been used in the analysis~\cite{Alt:2006jr}. Later on NA49 and NA61/SHINE experiments registered multiplicity distributions of negatively charged particles produced in p+p and the most central (1\%) Be+Be, Ar+Sc and Pb+Pb collisions at the same center of mass energy, but emitted to the full forward hemisphere, $y_{\pi}>0$~\cite{Motornenko:2017klp,Grebieszkow:2017gqx}. 

In this paper we focus on description of centrality dependence of average multiplicity and scaled variance of multiplicity distribution of charged particles produced in Pb+Pb collisions in $1.1 < y_{\pi} < 2.6$ as measured by the NA49 Collaboration. We also try to describe data on multiplicity fluctuations in the full forward hemisphere obtained in p+p interactions and the most central (1\%) Be+Be, Ar+Sc and Pb+Pb collisions.

\section{Model description}
\label{sect:model_desc}

We define cluster as a quasi-neutral gas of charged and neutral particles which exhibits collective behavior. The characteristic space scale of this shielding is the Debye length (or radius):
\begin{equation}
\lambda^{2}_{D}=\frac{kT}{4\pi e^{2}n}
\label{eq:Debye_length}
\end{equation}
where $n$ is the density of charged pions.
Taking pion radius $r_{\pi}=0.7$~fm~\cite{Bernard:2000qz} and $kT=0.15$~GeV, we have $n=0.46$~fm${}^{-3}$. Consequently, the Debye length is equal to $\lambda_{D}=4.2$~fm. In the Debye sphere of the volume
\begin{equation}
V=\frac{4}{3}\pi\lambda^{3}_{D}
\label{eq:Debye_volume}
\end{equation}
we have $N\simeq 143$ charged pions, what correspond at $\sqrt{s_{NN}}=17$~GeV to the number of projectile participants $N_{p}\simeq 18$. 

The statistical hadronization model is a very efficient tool for description of average particle multiplicities in high energy heavy ion reactions~\cite{Cleymans:1992zc,Yen:1997rv,Becattini:1997ii,BraunMunzinger:2001ip,Becattini:2003wp,BraunMunzinger:2003zd} as well as in elementary particle reactions~\cite{Becattini:1995if,Becattini:1997rv,Becattini:2001fg}. Within this model there is also possible to obtain multiplicity fluctuations since the status of the hadronizing sources is known. Multiplicity and electric charge fluctuations have been proposed as a good selective tool between hadron gas and quark-gluon plasma~\cite{Jeon:2000wg,Asakawa:2000wh} provided they survive the phase transition and the hadronic system freezes out in a nonequilibrium situation. To properly assess the selective power of such observables, one should first calculate fluctuations in a hadron gas by including effects of quantum statistics, conservation constrains, etc. The effects of conservation constrains on fluctuations in thermal ensembles were first addressed from the perspective of heavy ion collisions in Ref.~\cite{Stephanov:1999zu}. More recently, it has been pointed out~\cite{Begun:2004gs,Begun:2004zb} that in the canonical ensemble (CE) with exact conservation of charges, scaled variance of the multiplicity distribution of any particle does not converge to the corresponding grand canonical (GCE) value even in the thermodynamic limit, unlike the mean~\cite{Cleymans:1997ib,Keranen:2001pr}.

\begin{figure}
\begin{center}
\includegraphics[angle=0,width=0.48\textwidth]{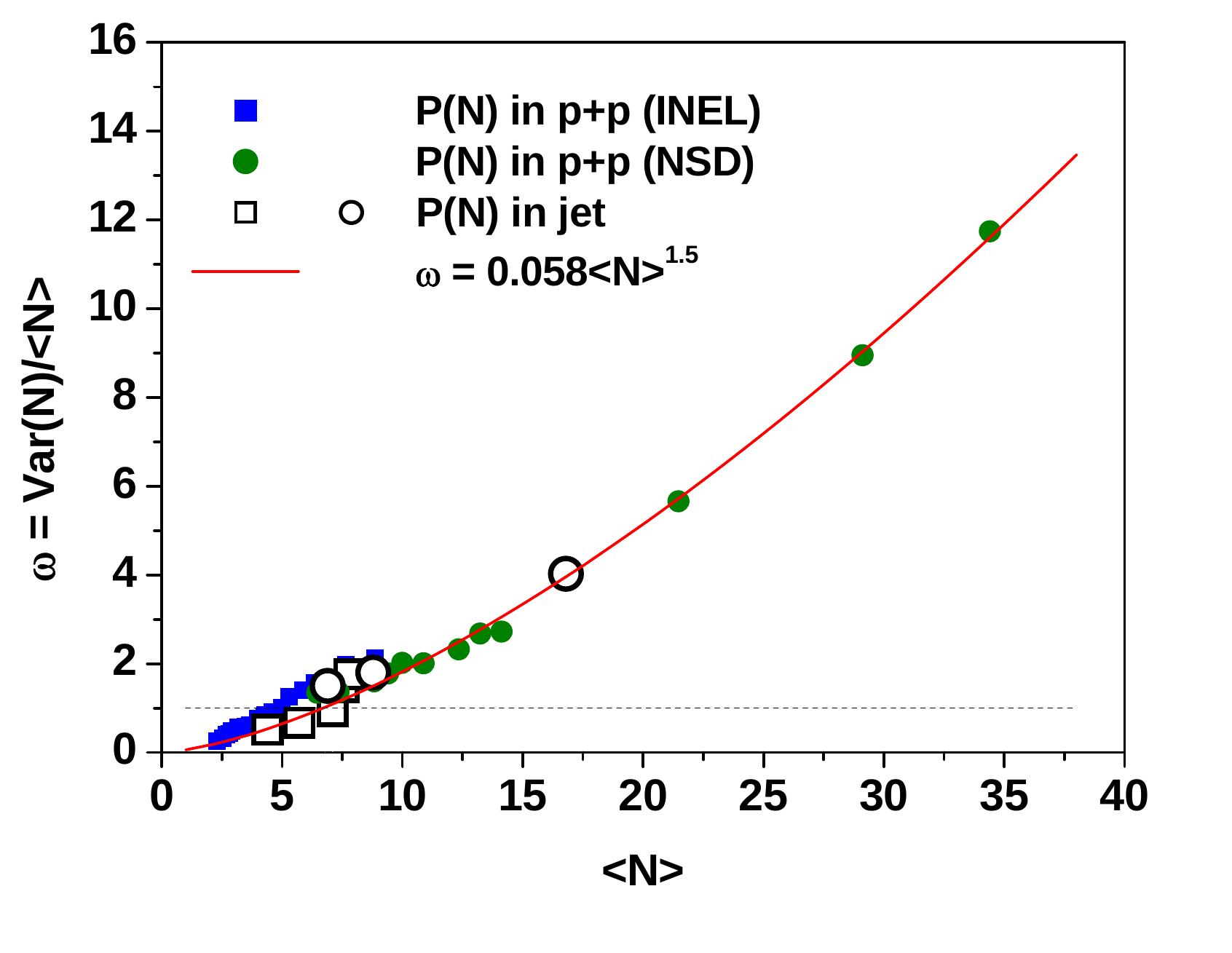}
\end{center}
\caption{\small Scaled variance of charged particles multiplicity distribution as a function of average charged multiplicity. By squares and circles we indicate data on particle production in p+p collisions: squares (inelastic data) are from the compilation for beam energy 3.7-303 GeV presented in~\cite{Wroblewski:1973tn}, full circles (non-single diffractive data) are from the compilation in~\cite{GeichGimbel:1987xy}. Open symbols are from data on particle production in jets: open circles are from~\cite{Aad:2016oit} and open squares from~\cite{Aad:2010ac,Aad:2011gn}. Line shows our fit to the data.}
\label{fig:varn_n}
\end{figure}

\begin{figure}
\begin{center}
\includegraphics[angle=0,width=0.48\textwidth]{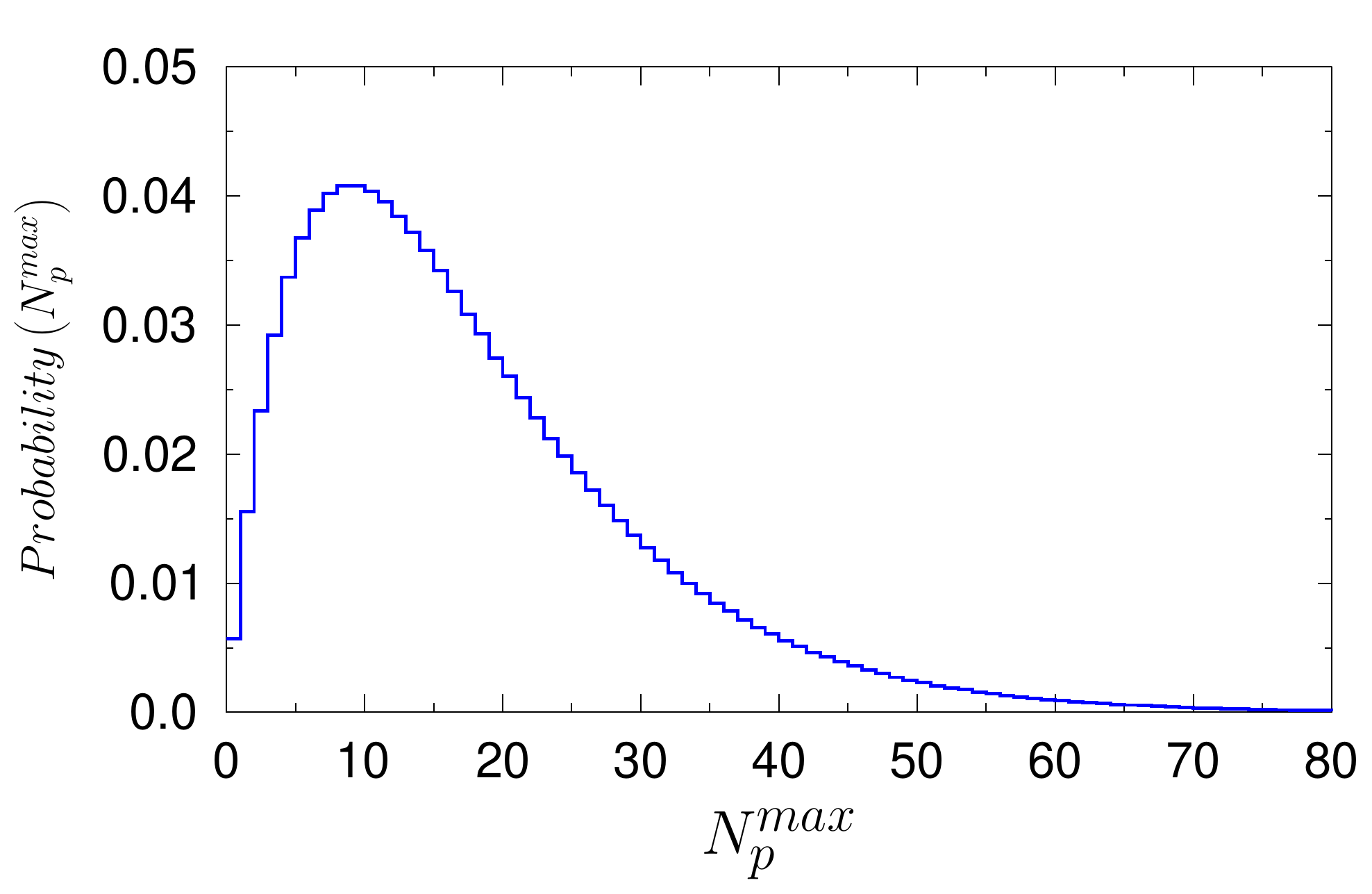}
\end{center}
\caption{\small Distribution of the number of nucleons which emit particles to the cluster, $N_{p}^{max}$. See text for details.}
\label{fig:npmax}
\end{figure}

If we split a CE, or micro-canonical ensemble (MCE) into N subsystems, the variance of any particle multiplicity distribution is not additive, as conservation constrains involve nonvanishing correlations between different subsystems even for large N. Thus, their GCE and CE thermodynamic limits differ. We split a CE with a large volume into cluster, which is a Grand Canonical Ensemble with the rest of the system being a reservoir~\cite{Becattini:2005cc}. Multiplicity distribution in a cluster is given by Negative Binomial distribution (NBD) while the rest (reservoir), treated as a superposition of elementary collisions, is described by Binomial distribution (BD). Variance of multiplicity distribution, $Var\left(N\right)$ depends on the mean multiplicity $\langle N\rangle$ of the system. For example, $Var\left(N\right)=\langle N\rangle\left(1+\sigma\langle N\rangle\right)$ with $\sigma=+1, -1, 0$ for Bose-Einstein, Fermi-Dirac and Boltzmann-Gibbs statistics, respectively. If $\langle N\rangle$ increase with energy $Var\left(N\right)$ also changes.

Fig.~\ref{fig:varn_n} presents a compilation of values of scaled variances of charged particle multiplicity distributions as a function of average charged multiplicity. Such dependence may be well fitted by a simple formula~\footnote{The rough formula (\ref{eq:varn_n}) asserts Taylor's law, $Var\left(N\right)=a\cdot \langle N\rangle^{b}$ with exponent $b>2$. Such behaviour corresponds a geometrical random walk (as opposed to the ordinary additive random walk) if multiplicity density at each step grows on average (super-critical model)~\cite{Cohen:2013}.}:
\begin{equation}
\omega\left(N\right)=0.058\cdot \langle N\rangle^{1.5}.
\label{eq:varn_n}
\end{equation}

Comparing multiplicity fluctuations in jets and in minimum bias proton-proton interactions one observes a kind of self-similarity of the multiparticle production processes~\cite{Wilk:2013jsa}. Regardless of the amount of the available energy, the variance is the same power function of the average multiplicity.

\begin{figure*}
\begin{center}
\includegraphics[angle=0,width=0.48\textwidth]{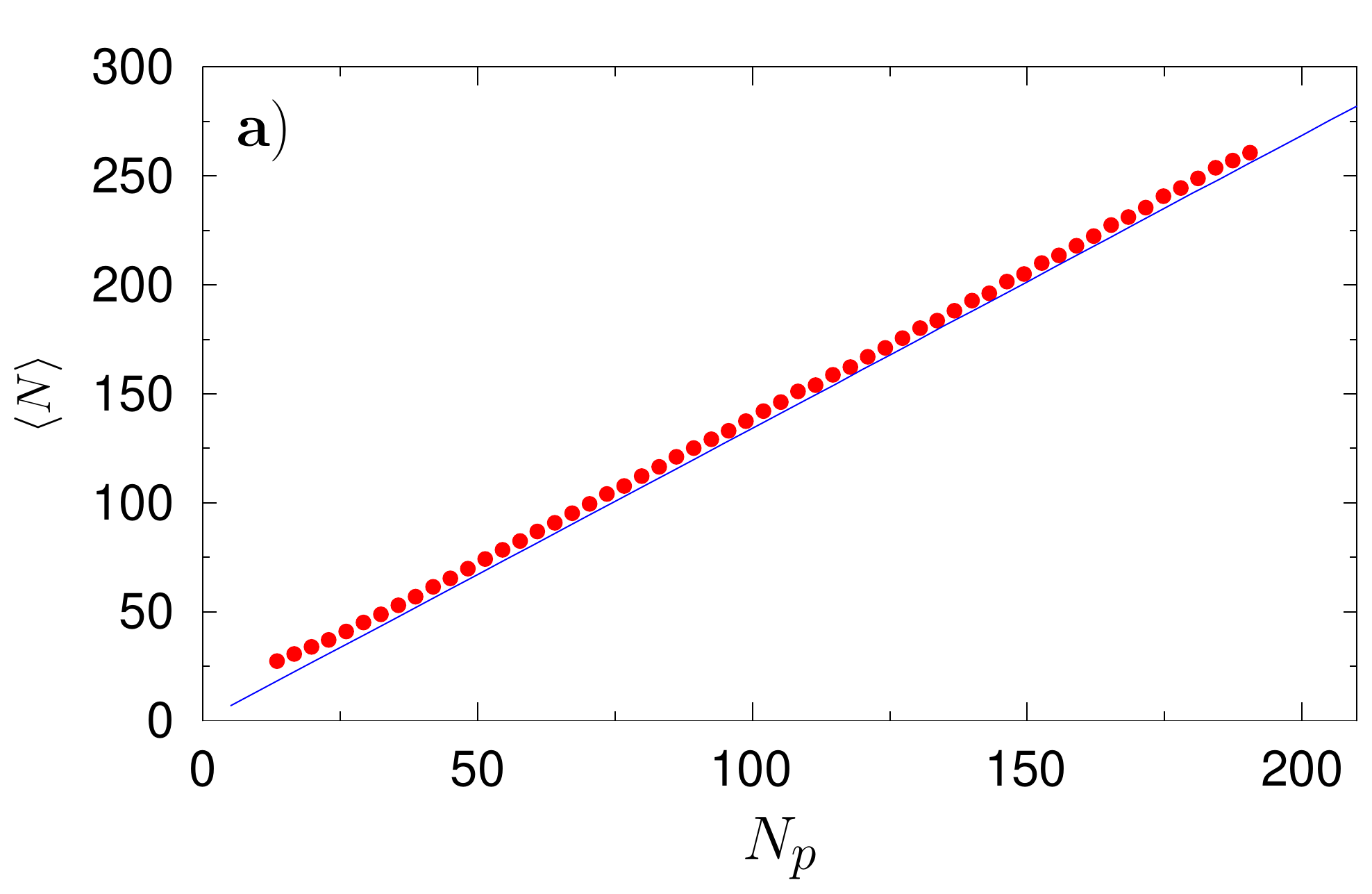}
\includegraphics[angle=0,width=0.48\textwidth]{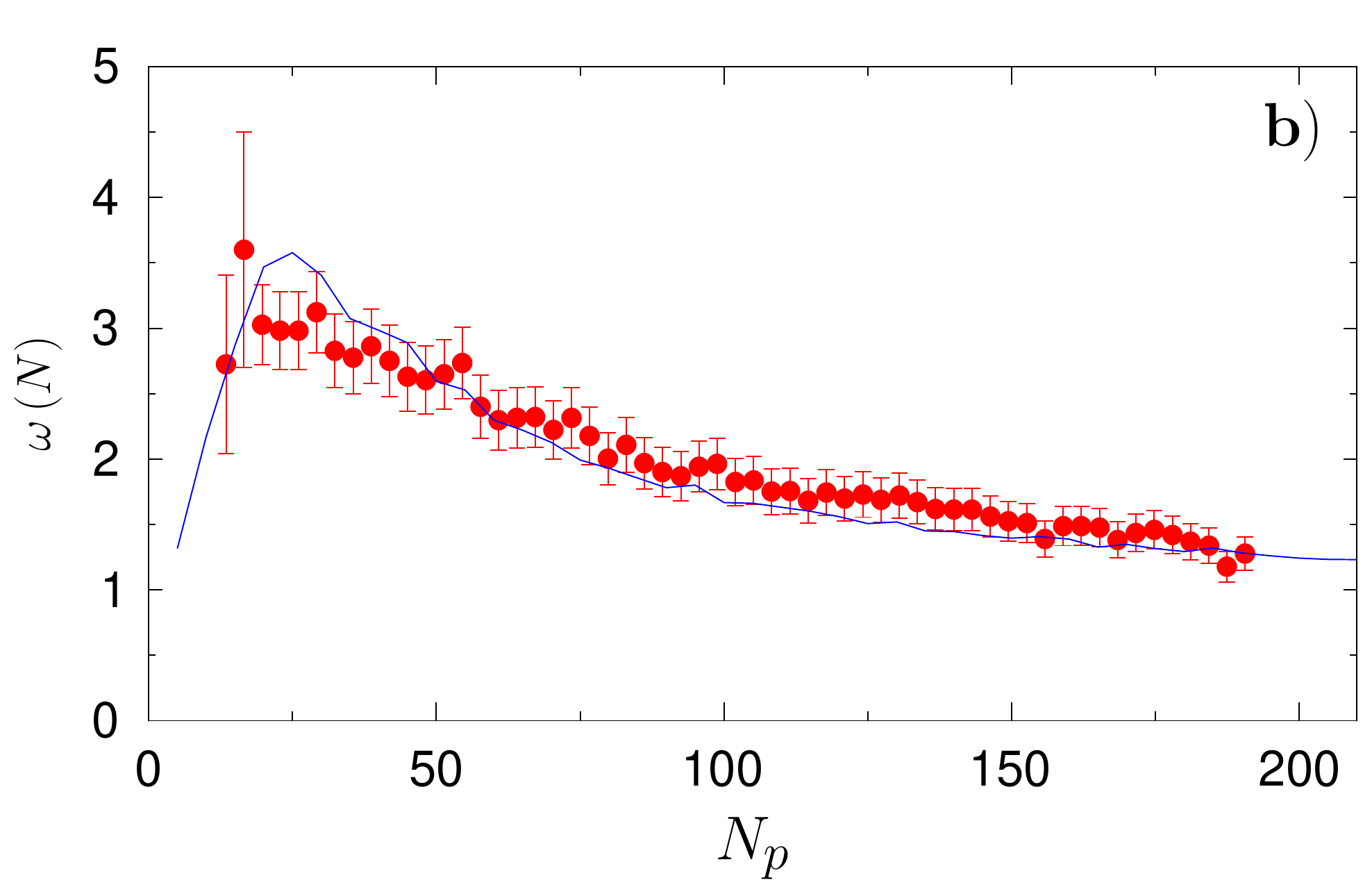}
\end{center}
\caption{\small Average number of all charged particles (panel a)) and scaled variance of all charged multiplicity distribution (panel b)) of particles produced in Pb+Pb collisions plotted as a function of number of nucleons from projectile nucleus which participate in the collision. Circles – NA49 data.}
\label{fig:fluct_all}
\end{figure*}

\begin{figure*}
\begin{center}
\includegraphics[angle=0,width=0.48\textwidth]{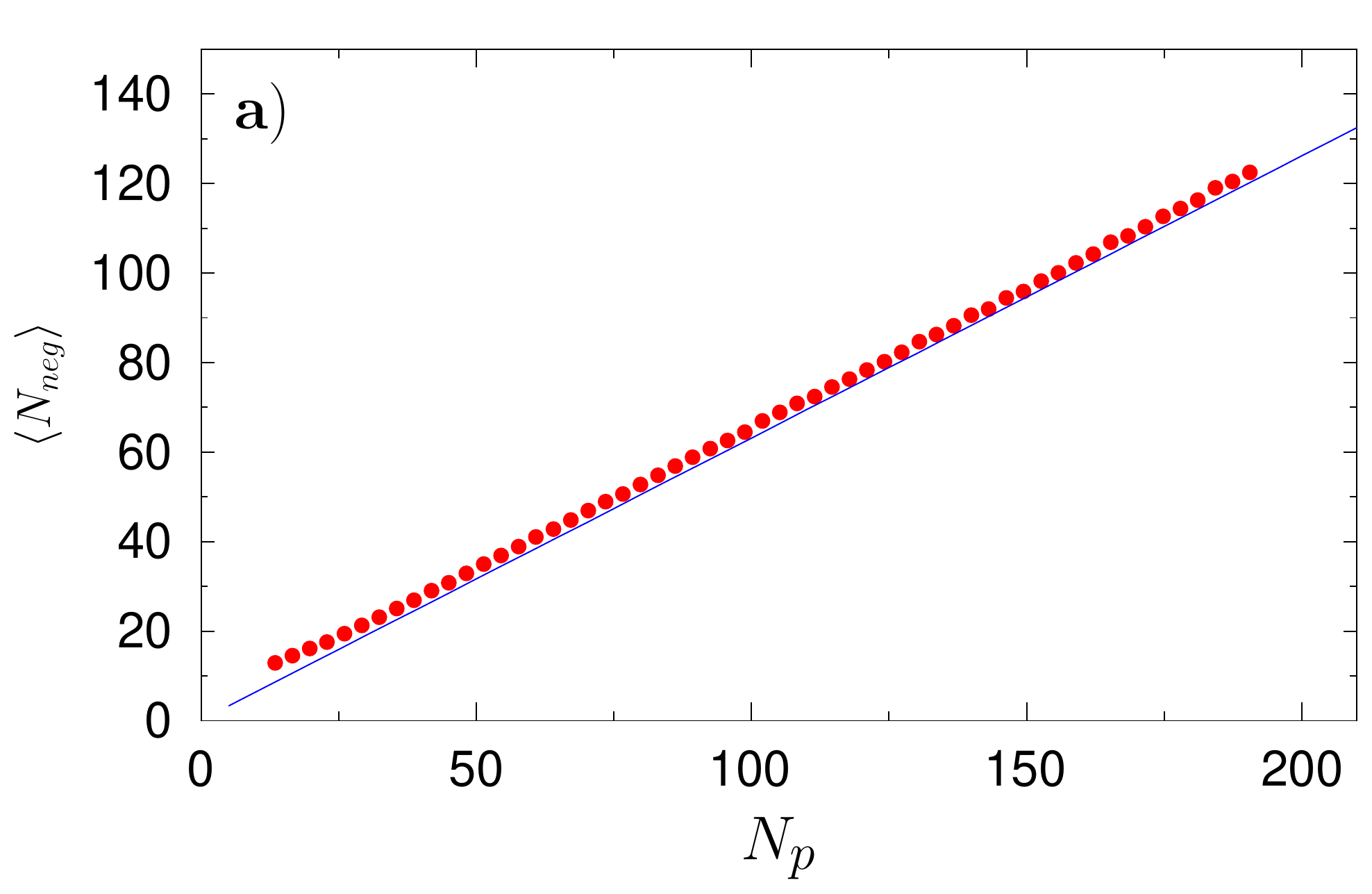}
\includegraphics[angle=0,width=0.48\textwidth]{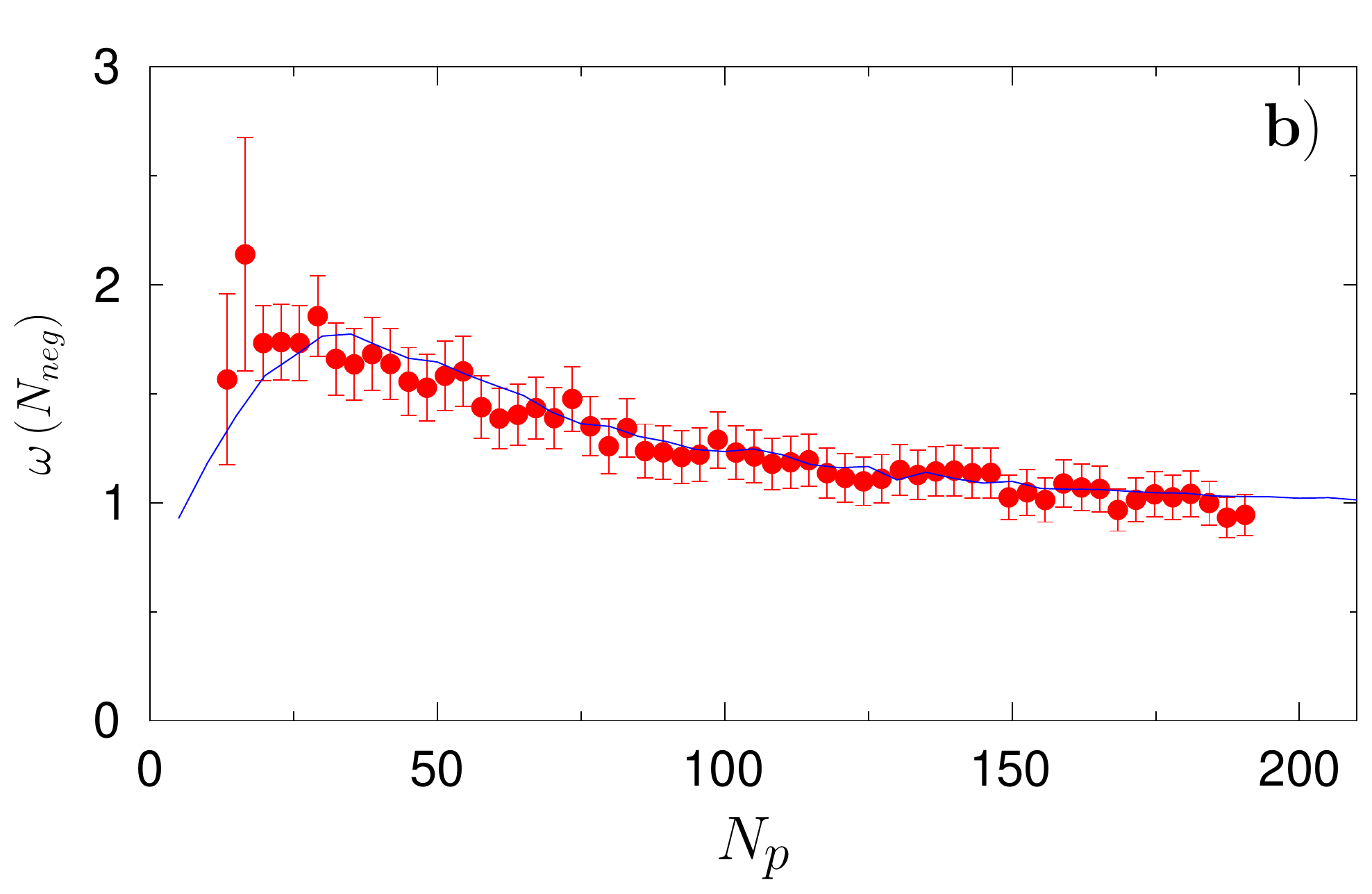}
\end{center}
\caption{\small The same as in Fig.~\ref{fig:fluct_all} but for negatively charged particles.}
\label{fig:fluct_neg}
\end{figure*}
\begin{figure}
\begin{center}
\includegraphics[angle=0,width=0.48\textwidth]{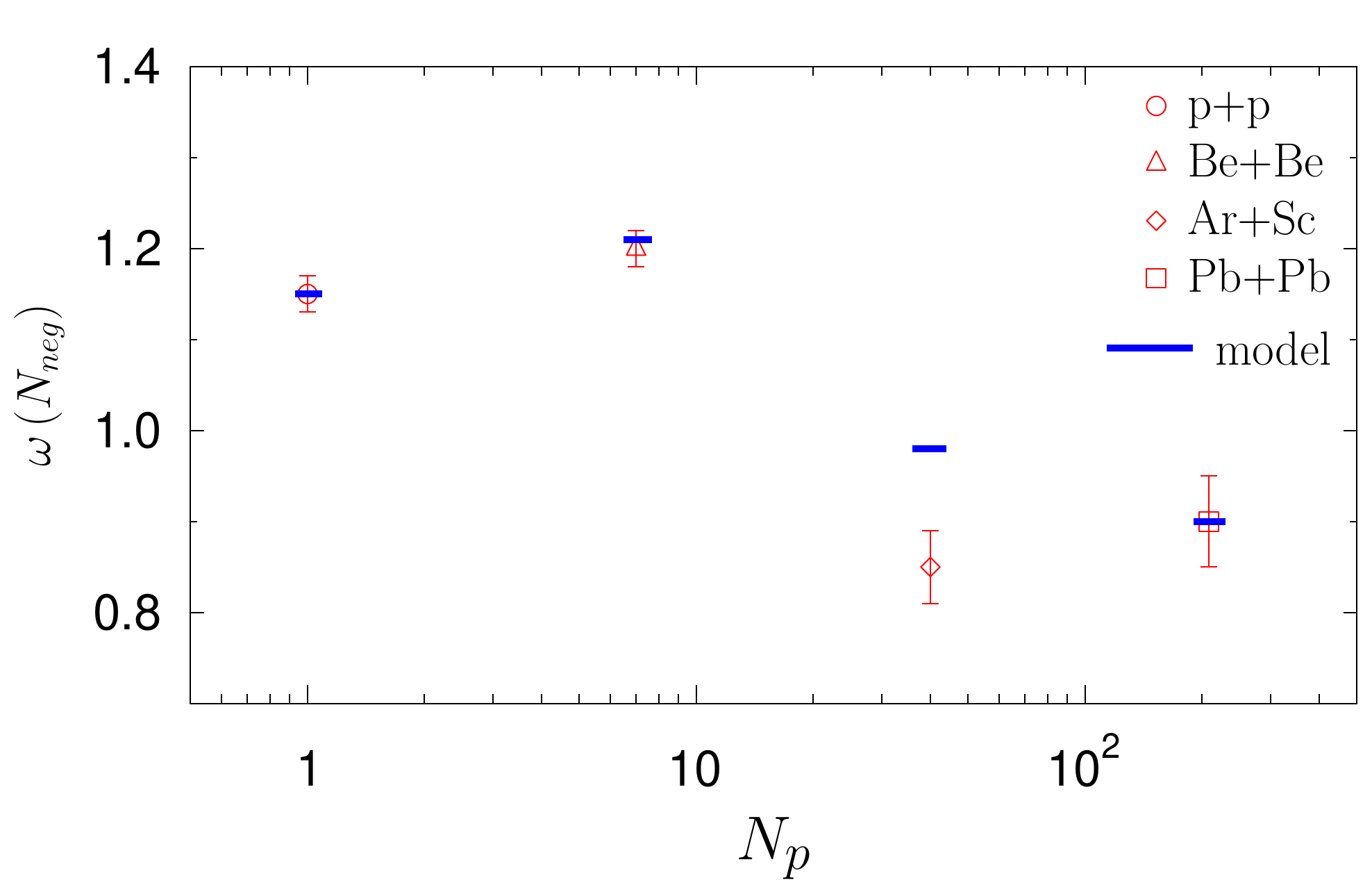}
\end{center}
\caption{\small Scaled variance of negatively charged multiplicity distribution of particles produced in p+p and the most central (1\%) Be+Be, Ar+Sc and Pb+Pb collisions, and emitted to the forward hemisphere, $y_{\pi}>0$ plotted as a function of number of nucleons from projectile nucleus which participate in the collision. Symbols present data of the NA49 and NA61/SHINE experiments~\cite{Motornenko:2017klp,Grebieszkow:2017gqx}. With the line we show values obtained using our model.}
\label{fig:fluct_na61}
\end{figure}

Negative Binomial distribution is a statistical tool commonly used for description of multiplicity distributions of particle produced in high-energy nuclear collision:
\begin{equation}
P_{NBD}\left(N,\langle N\rangle,k\right)=\binom{N+k-1}{N}\left(\frac{\langle N\rangle}{k}\right)^{N}\left(1+\frac{\langle N\rangle}{k}\right)^{-N-k}.
\label{eq:nbd_def}
\end{equation}
NBD has two free parameters: $\langle N\rangle$ describing mean multiplicity and, not necessarily integer parameter $k$ ($k\geq 1$) affecting shape of the distribution. Variance of NBD is given by:
\begin{equation}
Var\left(N\right)=\langle N\rangle+\frac{\langle N\rangle^{2}}{k}.
\label{eq:var_nbd}
\end{equation}
Both $\langle N\rangle$ and $k$ depend on collision energy. The energy dependence of average multiplicity of charged particles produced in proton-proton interactions may be well parameterized by~\cite{GeichGimbel:1987xy}:
\begin{equation}
\langle N_{ch}\rangle=A + B\ln s + C\ln^{2} s.
\label{eq:nch_param}
\end{equation}
where $\sqrt{s}$ is the center of mass energy of two colliding protons, and, $A=2.7\pm 0.7$, $B=-0.03\pm 0.21$, and $C=0.167\pm 0.016$. 
Parameterization~(\ref{eq:nch_param}) is valid for $\sqrt{s}$ ranging between 10 and 900 GeV. 
In proton-proton collisions the energy dependence of the NBD shape parameter $k$ is given by~\cite{GeichGimbel:1987xy}:
\begin{equation}
k^{-1}=a + b\ln \sqrt{s}.
\label{eq:k_param}
\end{equation}
with $a=-0.104\pm 0.004$, $b=0.058\pm 0.001$, and $s$ in $GeV^2$.
Using Eqs.~(\ref{eq:nch_param}) and~(\ref{eq:k_param}) one can obtain a NBD shape parameter $k$ as a function of average charged multiplicity, $\langle N_{ch}\rangle$:
\begin{eqnarray}
\lefteqn{k^{-1}\left(\langle N_{ch}\rangle\right)}\nonumber\\
 & = & 
 -0.104+0.0868\left(0.03+\sqrt{-1.8+0.668\cdot \langle N_{ch}\rangle}\right).
\label{eq:k_vs_nch}
\end{eqnarray}
Using Eq.~(\ref{eq:nch_param}) one may also find that at the interesting center of mass energy, $\sqrt{s}=17.3$ GeV:
\begin{equation}
\langle N_{ch}\rangle\left(\sqrt{s}=17.3 {\rm GeV}\right)=7.95.
\label{eq:av_nch}
\end{equation}

To describe the NA49 data the following particle clusterization method was used. Each projectile nucleon participating in collision “produces” particles independently,
\begin{equation}
\langle N\rangle=N_p\cdot \langle N_{ch}\rangle,
\label{eq:nch}
\end{equation}
where $\langle N\rangle$ is the average multiplicity produced in Pb+Pb collisions at particular centrality, $N_{p}$ is the number of nucleons from projectile nucleus participating in collision and $\langle N_{ch}\rangle$ is the average multiplicity produced in proton-proton interactions. Having calculated $\langle N\rangle$, the multiplicity in a given event of collision is calculated according to NBD distribution with the shape parameter $k$ dependent on $\langle N\rangle$, according to Eq.~(\ref{eq:k_vs_nch}). This is up to certain value of 
$N_{p}=N_{p}^{max}$, for which clusters of secondary particles may be formed. The value of $N_{p}^{max}$ is sampled from a Gamma distribution with $\langle N_{p}^{max}\rangle=18$ and $Var\left(N_{p}^{max}\right)=9\cdot\langle N_{p}^{max}\rangle$, see Fig.~\ref{fig:npmax}. The rest of colliding projectile nucleons, $m=N_{p}-N_{p}^{max}$ do not contribute their produced particles to the cluster. The produced by them particles are emitted according to Binomial distribution:
\begin{equation}
P_{BD}\left(N, n, p\right)=\binom{n}{N} p^{N}\left(1-p\right)^{n-N}.
\label{eq:bd_def}
\end{equation}
with $\langle N\rangle=\langle N_{ch}\rangle$, and probability $p=p_{BD}=0.4$. Clusters of particles are formed with a certain probability, $p_c=0.25$. If cluster of particles is not formed then all colliding nucleons emit their produced particles according to Binomial distribution.

\section{Results}
\label{sect:results}

The resultant centrality dependencies of average all charged multiplicities and corresponding scaled variances of multiplicity distributions are presented in Fig.~\ref{fig:fluct_all}. To include experimental acceptance we accepted a fraction of 17\% of generated particles, see Appendix for detailed discussion of acceptance.

Fig.~\ref{fig:fluct_neg} shows similar results as Fig.~\ref{fig:fluct_all} but for negatively charged particles. To obtain corresponding fits we had to adjust only one parameter: the average multiplicity in proton-proton collisions. For the case of negatively charged particles $\langle N_{ch}\rangle=3.6$. In the clustering model the only difference between negatively and all charged particles is the experimental acceptance manifested by the fraction of accepted particles.
Using similar considerations we have obtained the values for scaled variance of negatively charged multiplicity distribution produced in p+p and the most central (1\%) Be+Be, Ar+Sc and Pb+Pb collisions, and emitted to the forward hemisphere, $y_{\pi}>0$~\cite{Motornenko:2017klp,Grebieszkow:2017gqx}, see Fig.~\ref{fig:fluct_na61}.

\section{Concluding remarks}
\label{sect:cr}

In this paper we used the concept of clusterization in the mechanism of multiparticle production for description of multiplicity fluctuations observed in relativistic ion collisions at CERN SPS. Our results are as follows:
\begin{itemize}

\item 
It is shown that the observed non-monotonic behaviour of the scaled variance of multiplicity distribution as a function of collision centrality (such effect is not observed in a widely used string-hadronic models of nuclear collisions) can be fully explained by the correlations between produced particles promoting cluster formation.

\item
We defined a cluster as a quasi-neutral gas of charged and neutral particles which exhibits collective behaviour. The characteristic space scale of this shielding is the Debye length. 

\item
We split a Canonical Ensemble or a Micro Canonical Ensemble with a very large volume into cluster, which is by definition, a Grand Canonical Ensemble, with the rest of the system acting as a reservoir. Multiplicity distribution in a cluster is given by Negative Binomial distribution while the rest (reservoir), treated as a superposition of elementary collisions, is described by Binomial distribution. 

\item 
The ability to generate spatial structures (cluster phase) sign the propensity to self-organize of hadronic matter. Multiplicity clustering provide new insights on non-monotonic behaviour of multiplicity fluctuations.

\end{itemize}


\begin{acknowledgements}
The numerical simulations were carried out in laboratories created under the project
``Development of research base of specialized laboratories of public universities in Swietokrzyskie region'',
POIG 02.2.00-26-023/08, 19 May 2009.\\
MR was supported by the Polish National Science Centre (NCN) grant 2016/23/B/ST2/00692.
\end{acknowledgements}

\appendix

\section{Imprints of acceptance}\label{app:1}

Let us assume that $g\left(M\right)$ presents a real distribution which describe multiplicity distribution in the full phase space. Scaled variance $\omega$ is given by parameters of such distribution. For example:
\begin{equation}
\omega = 
\begin{cases} 
  1+\langle M\rangle / k & \text{for NBD}  \\
  1       & \text{for PD } \\
  1-\langle M\rangle /k & \text{for BD}
\end{cases}
\label{eq:real_omega}
\end{equation}
However, in the experiment we measure the multiplicity only within some window in rapidity, $\Delta y$. Roughly, for a fixed acceptance $\alpha<1$ we have 
\begin{equation}
\omega= \alpha\cdot \omega_{\alpha=1},
\label{eq:omega_acc}
\end{equation}
and scaled variance decrease monotonically with decreasing acceptance. Of course, such procedure is not correct. Let us assume that the detection process is a Bernoulli process described by the BD with the generating function
\begin{equation}
F\left(z\right)=1-\alpha+\alpha\cdot z,
\label{eq:BG_gen}
\end{equation}
where $\alpha$ denotes the probability of the detection of a particle in the rapidity window. The number of the registered particles is
\begin{equation}
N=\sum_{i=1}^{M} n_{i},
\end{equation}
where $n_{i}$ follows the BD with the generating function $F\left(z\right)$ and $M$ comes from $g\left(M\right)$ with the generating function $G\left(z\right)$. The measured multiplicity distribution $P\left(N\right)$ is therefore given by the generating function
\begin{equation}
H\left(z\right)=G\left[F\left(z\right)\right]
\end{equation}
and finally we have
\begin{equation}
P\left(N\right)=\frac{1}{N!}\frac{d^{N}H\left(z\right)}{dz^{N}}\Biggl\vert_{z=0}
\end{equation}
Note that such procedure applied to NBD, PD or BD gives again the same distributions but with modified parameters. The scaled variance is given by:
\begin{equation}
\omega = 
\begin{cases} 
  1+\alpha\langle M\rangle / k = 1+\langle N\rangle / k& \text{for NBD}  \\
  1       & \text{for PD } \\
  1-\alpha\langle M\rangle /k = 1-\langle N\rangle /k & \text{for BD}
\end{cases}
\label{eq:real_omega_acc}
\end{equation}
For all mentioned in Eq.~(\ref{eq:real_omega_acc}) distributions $\omega\rightarrow 1$ when $\alpha\rightarrow 0$. In the case of small acceptance, the observed $P\left(N\right)$ tends to PD.

The above discussed procedure is also very rough, because it neglects conservation constrains, eg. energy conservation. To investigate this effect we adopt a induced partition scenario for particle distribution in phase space~\cite{Rybczynski:2018ino}. Namely, if the available energy, which may be distributed among secondary particles $U=const$ is limited then we have the following conditional probability for the single-particle energy distribution~\cite{Rybczynski:2018ino}:
\begin{equation}
f\left(E|U\right)=\frac{f_{1}\left(E\right)\cdot f_{N-1}\left(U-E\right)}{f_{N}\left(U\right)}=\frac{N-1}{U}\left(1-\frac{E}{U}\right)^{N-2}.
\label{eq:cond}
\end{equation}
In the induced partition mechanism $N-1$ randomly chosen independent points $\{U_{1},\,\ldots,\,U_{N-1}\}$ split a segment $(0,U)$ into $N$ parts, whose length is distributed according to Eq.~(\ref{eq:cond}). The length of the k$th$ part corresponds to the value of energy $E_{k}=U_{k+1}-U_{k}$ (for ordered $U_{k}$). In our example it could correspond to the case of random breaks of a string in $N-1$ points in the energy space~\cite{Rybczynski:2018ino}.

\begin{figure}
\begin{center}
\includegraphics[angle=0,width=0.48\textwidth]{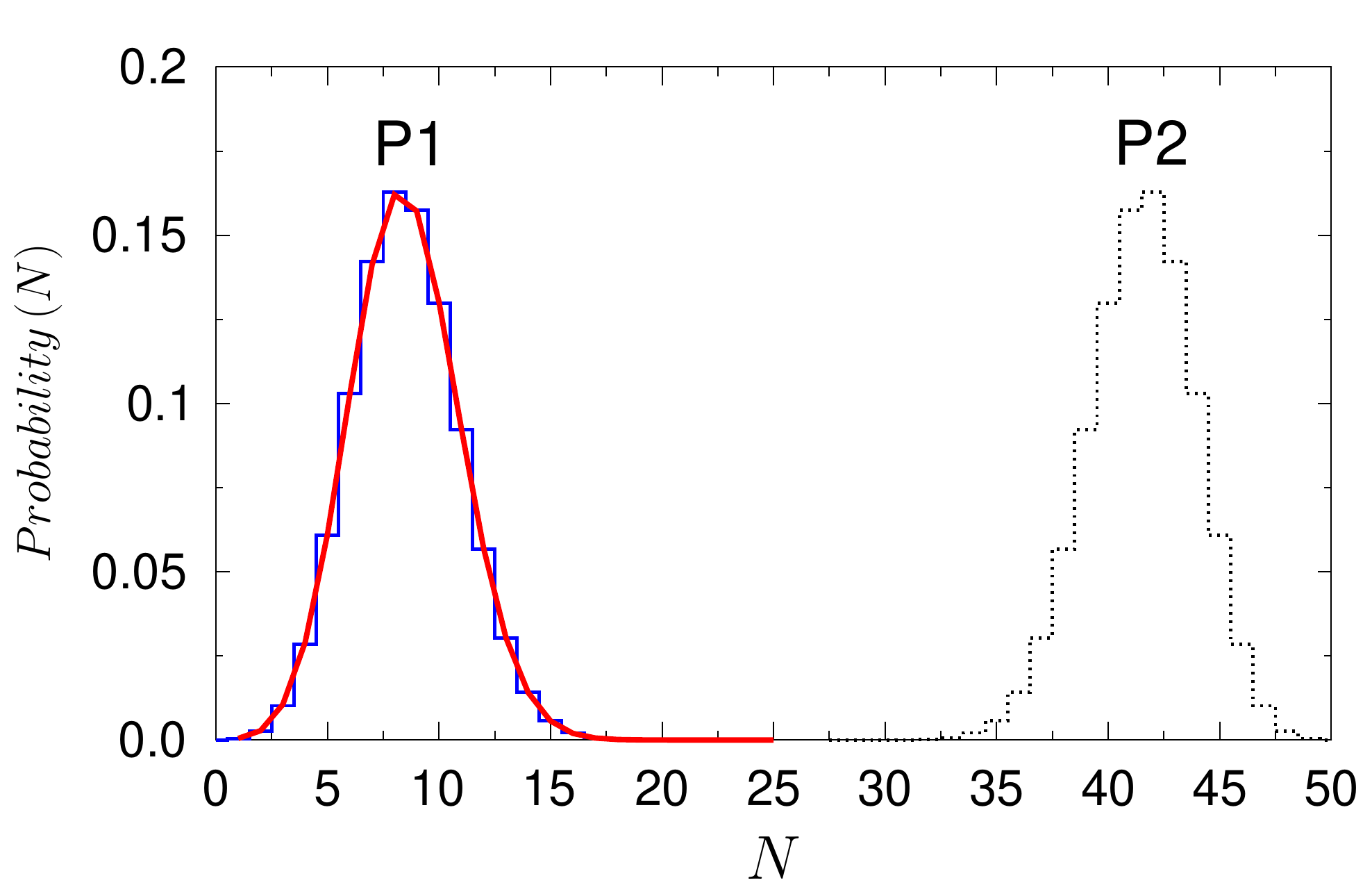}
\end{center}
\caption{\small Left histogram shows multiplicity distribution $P1\left(N\right)$ of 17\% least energetic particles from the original number of $50$, $U=100$~GeV. Red line show our BD fit. Right histogram presents multiplicity distribution $P2\left(N\right)$ of remaining 83\% of particles. See text for details.}
\label{fig:p12}
\end{figure}

\begin{figure}
\begin{center}
\includegraphics[angle=0,width=0.48\textwidth]{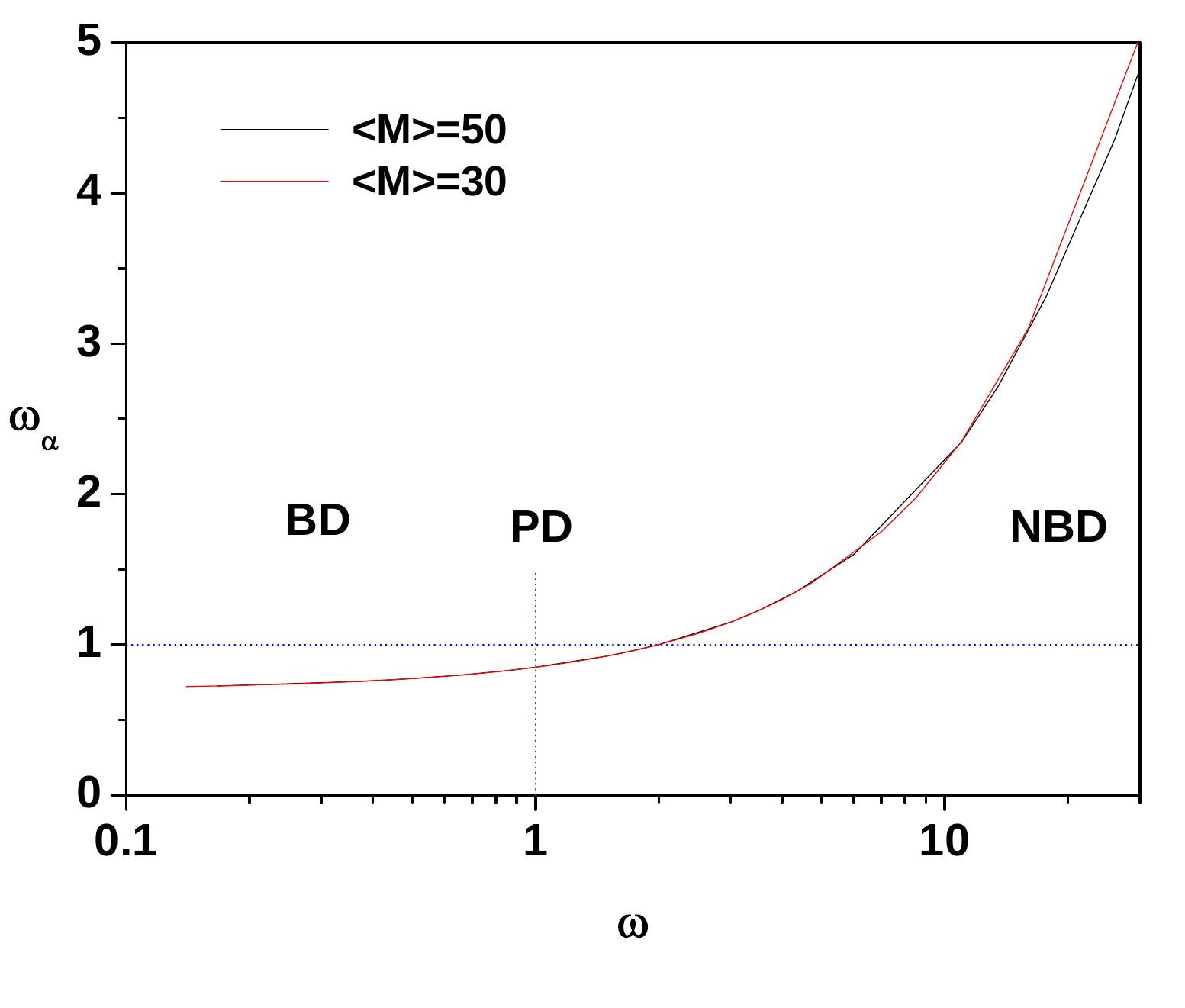}
\end{center}
\caption{\small $\omega_{\alpha}$ as a function of $\omega=\omega_{\alpha=1}$ for $P\left(M\right)$ given by BD, PD and NBD, with $\alpha=0.15$ and $z=2$. The results obtained for $\langle M\rangle=30$ and $\langle M\rangle=50$. See text for details.}
\label{fig:acc}
\end{figure}

To check the above considerations numerically let us take as an example a constant energy $U=100$~GeV and share it between a constant number, $N=50$ massless particles using the induced partition mechanism. Then we split the generated particles into two different multiplicity distributions, $P1\left(N\right)$ and $P2\left(N\right)$. If the energy of secondary particle is smaller than energy $E_{t}=0.38$~GeV then we put this particle into distribution $P1\left(N\right)$. Otherwise particle will populate distribution $P2\left(N\right)$. In such a way we put into multiplicity distribution $P1\left(N\right)$ about of 17\% of particles. In Fig.~\ref{fig:p12} we show both distributions, $P1\left(N\right)$ and $P2\left(N\right)$. Please note that both $P1\left(N\right)$ and $P2\left(N\right)$ multiplicity distributions have exactly the same variances. Moreover, the Pearson’s correlation coefficient calculated for distributions $P1\left(N\right)$ and $P2\left(N\right)$:
\begin{equation}
\rho\left(N_{P1},N_{P2}\right)=\frac{\langle N_{P1}\cdot N_{P2}\rangle-\langle N_{P1}\rangle\cdot \langle N_{P2}\rangle}{\sqrt{Var\left(N_{P1}\right)\cdot Var\left(N_{P2}\right)}}
\label{eq:p1p2_pear}
\end{equation}
equals $\rho\left(N_{P1},N_{P2}\right)=-1$. Multiplicity distribution $P1\left(N\right)$ may be easily fitted by BD, Eq.~(\ref{eq:bd_def}) with parameters $n=27.4$ and $p=0.31$, see Fig.~\ref{fig:p12}.

The ``measured'' multiplicity distribution is given by
\begin{equation}
P\left(N\right)=\sum_{M=N}^{\infty} P\left(M\right)\cdot P\left(N|M\right).
\label{eq:mult_meas}
\end{equation}
From the induced particle scenario we have the acceptance function:
\begin{equation}
P\left(N,M\right)=\frac{\Gamma\left(M/z+1\right)}{\Gamma\left(N+1\right)\cdot \Gamma\left(M/z-N+1\right)}\left(z\alpha\right)^{N}\left(1-z\alpha\right)^{M/z-N},
\label{eq:mult_ip}
\end{equation}
where $z$ is an parameter and $\alpha$ is the acceptance. 

As example in Fig.~\ref{fig:acc} we plot $\omega_{\alpha}$ as a function of $\omega=\omega_{\alpha=1}$ for $P\left(M\right)$ given by BD, PD and NBD, with $\alpha=0.15$ and $z=2$. The scaled variance $\omega_{\alpha}=1-\alpha$ for PD and reaches value $\omega_{\alpha}=1$ for NBD with $\omega=2$. For NBD we have $\omega_{\alpha} < \omega_{\alpha=1}$ and for BD $1-z\alpha\leq\omega_{\alpha} < 1-\alpha$. Such behavior is almost insensitive to $\langle M\rangle$, see Fig.~\ref{fig:acc}.


\end{document}